\documentstyle[preprint,eqsecnum,aps]{revtex}

\input epsf
\begin{document}
\title{Noise in Grover's Quantum Search Algorithm}
\author{B. Pablo-Norman and M. Ruiz-Altaba}
\address{Instituto de F\'{\i}sica\\
Universidad Nacional Aut\'onoma de M\'exico\\
A.P. 20-364, M\'{e}xico, D.F. 01000}
\maketitle

\begin{abstract}
Grover's quantum algorithm improves any classical search algorithm. We show
how random Gaussian noise at each step of the algorithm can be
modelled easily because of the exact recursion formulas available for computing the quantum amplitude in Grover's algorithm. We study the algorithm's intrinsic robustess when no quantum correction codes are used, and evaluate how much noise the algorithm can bear with, in terms of the size of the phone book and a desired probability of finding the correct result. The algorithm loses efficiency when noise is added, but does not slow down. We also study the maximal noise under which the iterated quantum algorithm is just as slow as the classical algorithm. In all cases, the width of the allowed noise scales with the size of the phone book as 
$ N^{-2/3}$.
\end{abstract}






\section{Introduction}

There exist problems where the algorithm that solves them scales
exponentially as the size of the input is increased, for example computing all possible chess games, factoring a very large number, etc. This dependence on the size makes them physically unsolvable
for large enough inputs. Quantum algorithms have been invented to bypass this problem, like Shor's 
\cite{Shor} that turns tractable the problem of factoring numbers, and
Grover's \cite{grover} that improves the classical search for an item in a phone book.  In fact, the classical search algorithm does not scale exponentially. Rather, it is linear in the size of the phone book;
Grover's quantum algorithm improves it to a square--root dependence.
Recently, an experimental application of a quantum algorithm has been
implemented \cite{Chuang}, and agreement between theory and experiment was
found.

Nevertheless, the strength of a quantum algorithm is also its weakness:
a quantum computer performs simultaneous operations over  large superpositions of states, which are very sensitive to decoherence. Fortunately, quantum correction codes have been
developed \cite{shor3,steane1} with which a quantum computer can recover
from errors in the presence of moderate decoherence. But these quantum
correction codes are subject themselves to decoherence, and it is not fully
understood how decoherence affects the correction itself. In this work,
we study the intrinsic robustness of Grover's algorithm, when quantum
correction codes are not implemented.

\section{Grover's quantum search algorithm}

Any classical algorithm for finding an item in a randomly ordered phone book
(whether deterministic or probabilistic) requires $N/2$ steps on the average, because the only way to perform the search is to analyze each item one by one until the searched--for item is found. Recently, Grover invented a quantum algorithm \cite{grover} that runs like $O\left( \sqrt{N}\right) $. Let us review it briefly.

In a phone book with $N=2^n$ entries, each item can be represented by a
binary label of length $n$ or, equivalently, by a pure state of $n$ spin $1/2 $ particles. The algorithm is based on constructing a coherent
superposition of all these states, and applying repeatedly certain unitary
transformations to it.

Assume, for concreteness, that the item we are looking for is represented by
the state $\mid \downarrow \downarrow \cdots \downarrow \rangle $, i.e. by $n $ spin--down particles.

The algorithm works via the repeated action of the unitary steps below,
starting from an initial state which we take to be the full coherent
superposition of all states in the system, namely 
\begin{equation}
\Psi_0=\frac 1{\sqrt{N}}\left( 
\begin{array}{c}
1 \\ 
1 \\ 
\vdots \\ 
1
\end{array}
\right) .  \label{1}
\end{equation}
Of course, one could start equally well with some other initial state 
\cite{lidar}.

The two unitary steps to be repeated are the following:

First, invert the phase of the looked--for state trough the unitary
transformation 
\begin{equation}
U_1=\left( 
\begin{array}{cccc}
-1 & 0 & \cdots & 0 \\ 
0 & 1 & 0 & \vdots \\ 
\vdots & 0 & \ddots & 0 \\ 
0 & 0 & \cdots & 1
\end{array}
\right) {\rm {.}}  \label{2}
\end{equation}

Secondly, invert, with respect to the average, the phase of the looked--for
state trough the unitary diffusion matrix 
\begin{equation}
\left( U_{2}\right) _{ij}=\frac{2}{N}-\delta _{ij}{\rm {.}}  \label{3}
\end{equation}
These two steps are equivalent to the action of the following single unitary
transformation: 
\begin{equation}
U=U_{2}U_{1}=\frac{2}{N}\left( 
\begin{array}{cccc}
-1+\frac{N}{2} & 1 & \cdots & 1 \\ 
-1 & 1-\frac{N}{2} & 1 & \vdots \\ 
\vdots & 1 & \ddots & 1 \\ 
-1 & 1 & 1 & 1-\frac{N}{2}
\end{array}
\right) .  \label{4}
\end{equation}

When the unitary transformation $U$ has been applied $m$ times to the
initial state $\Psi_0$, the new quantum state will be 
\begin{equation}
\Psi _m=U^m\Psi _0=\left( 
\begin{array}{c}
A_m \\ 
B_m \\ 
\vdots \\ 
B_m
\end{array}
\right) ,  \label{5}
\end{equation}
The action of $U$ on the initial state $\Psi_0$ yields only two distinct
amplitudes $A_m$ and $B_m$, whereby it is possible to recast the recursion
relation in just two dimensions. The restriction of $U$ to this
two--dimensional subspace will be denoted by $S$. Explicitly, the amplitudes 
$A_m$ and $B_m$ are given by the recursion formula 
\begin{equation}
\left( 
\begin{array}{c}
A_{m+1} \\ 
B_{m+1}
\end{array}
\right) =\left( 
\begin{array}{cc}
1-\frac 2N & 2-\frac 2N \\ 
\frac{-2}N & 1-\frac 2N
\end{array}
\right) \left( 
\begin{array}{c}
A_m \\ 
B_m
\end{array}
\right) =S\left( 
\begin{array}{c}
A_m \\ 
B_m
\end{array}
\right) =S^{m+1}\left( 
\begin{array}{c}
\frac 1{\sqrt{N}} \\ 
\frac 1{\sqrt{N}}
\end{array}
\right) ,  \label{6}
\end{equation}
The two--dimensional matrix $S$ has eigenvalues $e^{\pm i\varphi }$, with $\cos \varphi =1-\frac 1N$, whereby 
\begin{equation}
A_m=\frac 1{\sqrt{N}}\left( \cos m\varphi +\sqrt{N-1}\sin m\varphi \right)
\label{7}
\end{equation}
\begin{equation}
B_m=\frac 1{\sqrt{N}}\left( \cos m\varphi -\frac 1{\sqrt{N-1}}\sin m\varphi
\right)  \label{8}
\end{equation}

From (\ref{7}), the probability of finding the state we are looking for if
we measure $\Psi _{m}$ is thus 
\begin{equation}
P({m})=\left| A_{m}\right| ^{2}=\frac{1}{N}\left( \cos m\varphi +\sqrt{N-1} 
\sin m\varphi \right) ^{2}  \label{9}
\end{equation}
With the change of variables $\varphi =2\theta $, $P(m)$ can be written as 
\cite{boyer}: 
\begin{equation}
P({m})=\sin ^{2}\left( \theta \left( 2m+1\right) \right) {\rm {,}}
\label{10}
\end{equation}

Clearly, $P(m)$ is periodic, with maxima at 
\begin{equation}
\theta \left( 2m+1\right) =n\pi ,\qquad n{\rm {\ integer,}}  \label{11}
\end{equation}
The first maximum for large $N$ is approximately at 
\begin{equation}
m_{{\rm max}}\simeq \frac{\pi \sqrt{N}}4.  \label{12}
\end{equation}
and $P_{{\rm max}}=P(m_{{\rm max}})\simeq1$. The number of steps required to find the state with almost certainty scales like $\sqrt N$, as shown in (\ref{12}).

\section{Modelling Noise in Grover's Algorithm}

As stated in the introduction, quantum correction codes have been developed
and it is supposed that in the presence of low but physically realistic
levels of noise they are useful \cite{shor3,steane1}. These codes can be
implemented only if a small enough subset of the quantum computer's
q-bits undergo errors, and when the probability of occurrence of an error in
the computation is lower than a certain bound. On the other hand, the real
effect of the noise introduced by these correction codes over the original
algorithm is not completely known, because they are quantum computations
too. Hopefully, such errors are small and tractable. But what happens if
they are not? Or, even worse, what happens if many q-bits undergo errors? Is
it still possible to make sense of the computation under this hypothetical
noisy situation when quantum correction codes do not suffice or cannot be
implemented? If it does, how much noise the algorithm can bear with on its
own? We now turn to the answer to these questions.

In the particular case of Grover's algorithm, there is a simple way to model
 noise, because of the explicit recursion
formula (\ref{6}) for the amplitudes of the searched--for state.

Suppose that in each step of the algorithm, a white or Gaussian noise
modifies the state of the whole phone book according to 
\begin{equation}
\left( 
\begin{array}{c}
A_{m+1} \\ 
B_{m+1}
\end{array}
\right) =\frac{1}{Norm}\left[ S\left( 
\begin{array}{c}
A_{m} \\ 
B_{m}
\end{array}
\right) +\left( 
\begin{array}{c}
a_{m} \\ 
b_{m}
\end{array}
\right) \right] {\rm {,}}  \label{13}
\end{equation}
where $S$ is defined in (\ref{6}), and both $a_{m}$ and $b_{m}$ are noise, determined randomly by the
standard deviation $\sigma $ (common to both, for simplicity) of their
Gaussian distribution. Of course, the new state $\Psi _{m+1}$ is
appropriately normalized (that's what the denominator $Norm$ is for). Explicitly, 
\begin{equation}
\left( 
\begin{array}{c}
a_{m} \\ 
b_{m}
\end{array}
\right) =\sqrt{-2\sigma \log x_{1}}\left( 
\begin{array}{c}
  \sin 2\pi x_{2}  \\ 
\cos 2\pi x_{2} 
\end{array}
\right) \label{fla}
\end{equation}
where $x_1$ and $x_2$ are computer--generated random variables uniformly distributed over the interval $[0,1]$. The two Gaussian variables $a_{m}$ and $b_{m}$ are mutually independent,
and change, randomly, from one iteration of equation  (\ref{13}) to the next.  Note that when $\sigma=0$, $a_m$ and $b_m$ are always zero and thus there is no noise. 
 
A crucial {\sl caveat} is in order here: note that we introduce only two different errors, one for the
searched--for state and one for all the other pure states. This
approximation is physically unrealistic, but worthy of study. The full noisy
situation would call for allowing $N$ different random variables to be added
independently to each of the $N$ components of the state vector, instead of restricting ourselves to noise in the two--dimensional subspace where $S$ (instead of $U$) acts.

Now, we want to find the maximal allowed noise, quantified by  $\sigma $, in terms of both (a) the size $N$ of the phone book and (b) a given  probability $P_{{\rm cut}}$ for finding the searched-for state after a suitable number of iterations. If we set $P_{{\rm cut}} = P_{{\rm max}}$, then of course $\sigma$ can only be zero. As we allow for a decreased certainty of finding the result, and thus decrease $P_{{\rm cut}}$, the algorithm can bear with an increasing amount of noise. In the absurd limit of being happy with $P_{\rm cut}\simeq0$, which means we will not find the result, then any amount of noise is allowed. Of course, for any given $P_{{\rm cut}}>0$, a large enough noise will destroy the algorithm. In the next section we  establish the dependence of this maximal allowed noise, 
$\sigma_{\rm max}$, in terms of $N$ and $P_{\rm cut}$.

\subsection{Computations and Results}

To find when the algorithm breaks down as we increase the noise, we
treat the noise as a perturbation on the exact algorithm (recovered when  $\sigma =0$). Beforehand, we fix the phone book's size $N$ and the desired probability of finding the result, $P_{\rm cut}$. 

First, we take a very small initial value of $\sigma $ and evolve the initial state $\Psi_0$ in equation  (\ref{1}) according to the noisy iteration given in equation  (\ref{13}). After $m$ iterations, the probability $P(m)$ of finding the result is still $\mid A_m \mid^2$, where now the amplitude $A_m$ includes $m$ additions of noise.   It turns out that, on the average, $P(m)$ still reaches its maximum after $m_{{\rm max}}$
 steps. This is a pleasant surprise. At first thought, one could have imagined that noise not only decreased $P_{\rm max}$ (as it does), but also slowed down the algorithm (which it does not). To maximize the likelihood  of finding the result we must measure the quantum state after $m_{\rm max}$ iterations, with $m_{\rm max}$ given by the noiseless equation (\ref{12}).

Now we compute $P_{\max}=P(m_{\rm max})$ and compare it with $P_{\rm cut}$.
If $P_{{\rm max}}$ is greater than $P_{{\rm cut}}$~, we
increase the value of $\sigma $ and repeat the computation, otherwise we stop (see the Appendix for details). In this way, we find the maximal $\sigma$, labelled $\sigma _{{\rm max }}$, which is the limiting noise for   $P_{{\rm max}}\geq P_{{\rm cut}}$. Because of the probabilistic nature of the computations, we repeat this computation of $\sigma_{\rm max}$ many times (two-hundred): the value of $\sigma _{{\rm max }}$ we exhibit is the average, with a statistical error. 

We have carried out the evaluation of $\sigma_{\rm max}$ for seven 
different phone book sizes $N=2^{n} $ (with $n$ from $10$ to $16$) and for
five different values of $P_{{\rm cut}}$ (from $0.9$ to $0.5$ in steps of $0.1$).

For fixed $P_{{\rm cut}}$, the dependence of $\sigma _{\max }$ on $N$ is
always of the form: 
\begin{equation}
\sigma _{\max }\left( N,P_{{\rm cut}}\right) =\alpha \left( P_{{\rm   cut}}\right) N^{\phi }{\rm {,}}  \label{14}
\end{equation}
where $\phi $ is a true constant, found to be 
\begin{equation}
\phi =-0.696\pm 0.027  \label{xxy}
\end{equation}
and $\alpha $  varies smoothly from $.9$ to $.15$ as $P_{{\rm cut}}
$ decreases from $.9$ to $.5$ (see Appendix).

One of our main results is that the amount of noise that the algorithm can
handle decreases roughly as $N^{-2/3}$ with the size $N$ of the list. In
general, since the number of steps needed in each iteration is of the order
of $N^{1/2}$, and at each step we add a noise of width $\sigma$, we expect the maximal allowed $\sigma_{\max }$ to decrease
with $N$ faster than $N^{-1/2}$. Equivalently, we expect $\phi$ to be
smaller than minus one--half. The actual value found, equation (\ref{xxy}),
satisfies this bound. We have not found a general analytic argument to pin
down the actual value of $\phi$.

Alternatively, keeping $N$ fixed instead of $P_{cut}$, the relation between $\sigma_{{\rm max}} $ and $P_{{\rm cut}}$ can be written as 
\begin{equation}
\sigma _{\max }\left( N,P_{{\rm cut}} \right) =\gamma \left( N\right)
-\delta \left( N\right) P_{cut}  \label{15}
\end{equation}
where $\gamma $ goes from $0.0024$ to $0.00015,$ and $\delta $ from $-0.0020$ to $0.00013$ (${\rm log}_2N=n=10$ and $16,$ respectively), with errors of about $10\% $ (see Appendix for details). This means that the width of the maximal white noise that may be allowed increases linearly with decreasing  $P_{{\rm cut}}$.

Note that equations (\ref{14}) and (\ref{15}) are just convenient slices of a
surface in the three--dimensional space with co-ordinates 
$\left( N,P_{{\rm cut}}, \sigma _{{\rm max }}\right) $.

\section{Grover's algorithm is useful even if P$_{cut}<0.5.$}

In the derivation of the above results we exploited the experimental fact
that the number of steps needed to find the searched--for state does not
change when noise is present. Thus, another way to estimate the real maximal
noise that the noisy Grover's algorithm can handle, while still improving
the results of the classical search algorithm, is to let $P_{{\rm cut}}$ be
even lower than $0.5$. We now explain this.

Since $m_{{\rm max}}=\frac{\pi }{4}\sqrt{N}$ is always bigger than $N/2$,
there is an integer $I_N$ such that $I_N m_{{\rm max}}\leq {N}/{2}$,
namely 
\begin{equation}
I_N \simeq \frac{2}{\pi }\sqrt{N}  \label{16}
\end{equation}
Therefore, we can repeat the quantum search $I_N$ times with a low 
$P_{{\rm cut}}$ such that $1-\left( 1-P_{{\rm cut}}\right)^{I_N}\geq 0.5$. We are assured that we will find the searched--for state with probability
one-half in the same
number of steps as the classical algorithm. Of course, the classical algorithm find the result for sure, and compared with that finding the result only half the time is not very satisfactory. Instead of .5, we could equally well have chosen some other (higher) probability to be satisfied with, but we take .5 for definiteness as the extreme, illustrative case. The point is that the $P_{\rm cut}$ we need to enforce on the noisy quantum algorithm is smaller than .5. Note also that we are disregarding
the ${\rm log}_2 N$ steps needed in each of the $I_N$ independent iterations
to prepare the initial state $\Psi_0$. Including them would of course lower
a bit the maximal allowed noise.

The limiting probability at maximum with which the iterated quantum
algorithm is as slow as the classical one is 
\begin{equation}
P_{{\rm cut}}\geq 1-0.5^{\pi /(2\sqrt{N})}{\rm {,}}  \label{17}
\end{equation}
The meaning of this is, again, that we can let $P_{{\rm cut}}$ be smaller
than $0.5$ for a given $N$ because if we run $I_N$ times the quantum algorithm with $m_{{\rm max}}\simeq \pi \sqrt{N}/4$ steps, we will find the
searched--for state with a probability of at least $0.5$, and the total
number of steps will be less or equal to $N/2$ (ignoring the $\log _2N$
steps required for constructing the initial state $\Psi _0$).

To estimate this maximal noise that the quantum algorithm can bear before it
slows down all the way to equivalence with the classical one, we proceed as
follows. First, we choose the size $N$ of the list to be searched, and keep
it fixed. Then, using the bound (\ref{17}), we determine $P_{{\rm cut}}$, which is very low.
Finally, equation  (\ref{15}) yields $\sigma _{\max }$, which is now significantly higher. For a variety of $N$, our
results are shown in Table~1.

\begin{table}[tbh]
\begin{center}
\begin{tabular}{|c|c|c|c|}
\hline
$N$ & $P_{cut}$ & $\sigma _{\max }$ & $\Delta \sigma _{\max }$ \\ \hline
$1024$ & $0.034$ & $2.33\times 10^{-3}$ & $1.0\times 10^{-4}$ \\ \hline
$2048$ & $0.024$ & $1.48\times 10^{-4}$ & $4.3\times 10^{-5}$ \\ \hline
$4096$ & $0.017$ & $9.03\times 10^{-4}$ & $2.6\times 10^{-5}$ \\ \hline
$8192$ & $0.012$ & $5.68\times 10^{-4}$ & $1.6\times 10^{-5}$ \\ \hline
$16384$ & $0.0085$ & $3.28\times 10^{-4}$ & $1.7\times 10^{-5}$ \\ \hline
$32768$ & $0.0060$ & $2.13\times 10^{-4}$ & $1.7\times 10^{-5}$ \\ \hline
$65536$ & $0.0043$ & $1.17\times 10^{-4}$ & $1.1\times 10^{-5}$ \\ \hline
\end{tabular}
\end{center}
\caption{In the iterated quantum algorithm, for various sizes $N$ of a
phone book, the absolute maximal allowed Gaussian width $\sigma_{{\rm max}}$
of the white noise, and its statistical uncertainty (between $5$ and $10$~\%). Also shown is the (low!) limiting probability $P_{{\rm cut}}$ at
maximum.}
\end{table}

In figure 1, we plot $\sigma _{\max }$ as a function of $N$ for the data of
Table~1; the equation which fits it is 
\begin{equation}
\sigma _{{\rm max }}=\left( 0.275\pm 0.031\right) N^{(-0.68\pm 0.01)}{\rm  {,}}  \label{19}
\end{equation}
note that the exponent of $N$ in (\ref{19}) is essentially the same as the
exponent $\phi$ in (\ref{14}), even though $P_{{\rm cut}}$ depends on $N$
and is one or two orders of magnitude smaller than in Section~III.

\begin{figure}[tbp]
\let\picnaturalsize=N
\def\picsize{5.0in}
\def\picfilename{robfig1.eps}
\centerline{\ifx\picnaturalsize N\epsfxsize \picsize\fi \epsfbox{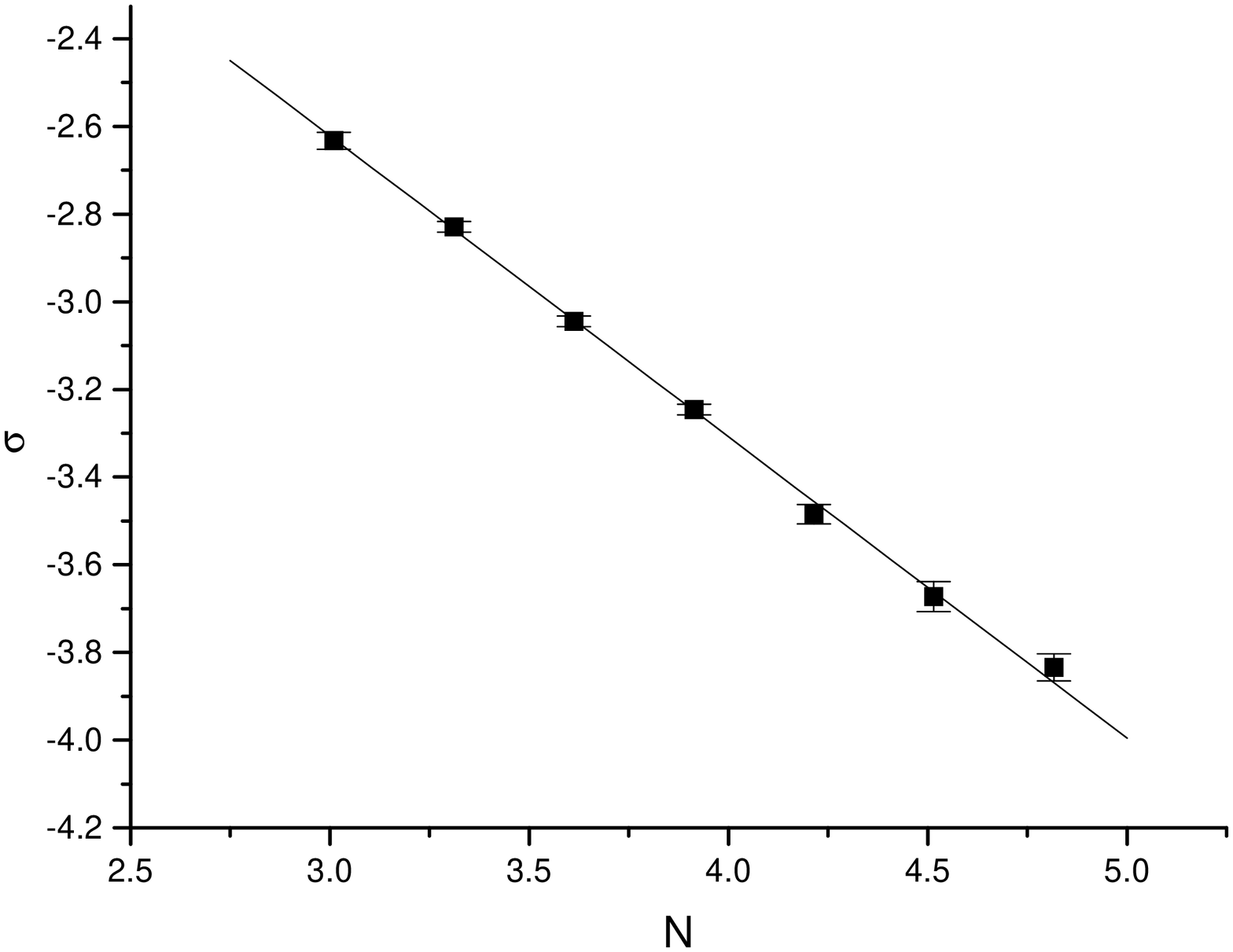}}
{\small Fig.1.  Plot of log $\sigma_{\rm max} $ as a function of  $\log N$ for the iterated quantum algorithm with  minimal $P_{\rm cut}= 1-0.5^{\pi /(2\sqrt{N})}$ that still improves the classical search algorithm. Even though to  each $N$ corresponds a different $P_{\rm cut}$, the plot  still displays the universal $N^{-2/3}$ dependence. }
\end{figure}

\section{Conclusions}

At the moment, quantum correction codes are restricted to the case when only
a enough small subset of the quantum computer's $q$-bits undergo errors, and
the probability of occurrence of an error is smaller than some bound, but it is believed that quantum computations
will be possible with physically realistic levels of noise even if the quantum correction codes employed undergo
errors themselves.

With this in mind, we studied the intrinsic robustness of Grover's
quantum search algorithm in a noisy environment. We modelled the noise with a single parameter, the width of a Gaussian distribution, and allowed for two independent noises at each step of Grover's quantum algorithm. 

We found that the quantum search algorithm still reaches the maximum
likelihood of finding the searched--for state in $\pi \sqrt{N}/4$ steps. The
strongest effect of noise is to decrease the maximum probability from
virtually $1$ (the noiseless case) to lower values, depending on the size of
the noise, equations (\ref{14}) and (\ref{15}). How much noise can we add to the
quantum computer, with the criterion that a repeated application of the
quantum algorithm is still faster than the classical one is given by equation (\ref{17}). In both cases, the allowed maximal noise decreases with the size of the phone book approximately as $N^{-2/3}$.

The presence of noise and the absence of quantum correction
codes is not completely disastrous: the quantum search algorithm can handle
by itself a reasonable amount of noise. Nevertheless, for large enough
databases, the allowed noise becomes tiny.

{\bf Acknowledgments.} This work is supported in part by CONACYT 25504-E
and DGAPA--UNAM IN103997. B.P.N. enjoys a scholarship from CONACYT.

\section{Appendix}

The computer program we used to derive the results in section III needs an
initial value of $\sigma $ (which we set to zero), and then computes $P({m})$.  If $P(m_{{\rm max }})>P_{{\rm cut}}$, then the
algorithm increases $\sigma$, repeating the process until the bound is
surpassed. This gives one value for $\sigma_{{\rm max}}$. We repeat the
whole story again and again and average over the values of $\sigma_{{\rm max}}$ found.

Let us illustrate our procedure with an example.

Let $P_{{\rm cut}}=0.7$. For each $n$ from $10$ to $16$, the program
increases the value of $\sigma $ starting from $0$ in steps of 
${\rm d}\sigma =0.0001$. The average maximal values of $\sigma$ thus found (in 200 runs) is then $\sigma _{{\rm max}}$, shown below with its statistical
uncertainty. Note that the error seems dominated by the step size: 
\begin{equation}
\begin{tabular}{ccc}
$N$ & $\sigma _{\max }$ &  \\ 
$1024$ & $0.0033 \pm 0.0007$ &  \\ 
$2048$ & $0.0022 \pm 0.0005$ &  \\ 
$4096$ & $0.0014 \pm 0.0004$ &  \\ 
$8192$ & $0.00098 \pm 0.00028$ &  \\ 
$16384$ & $0.00070 \pm 0.00020$ &  \\ 
$32768$ & $0.00048 \pm 0.00019$ &  \\ 
$65536$ & $0.00011 \pm 0.00017$ & 
\end{tabular}
\label{20}
\end{equation}

Taking a smaller step, ${\rm d}\sigma=0.00001$, we carry through the same
computations and find instead: 
\begin{equation}
\begin{tabular}{cc}
$N$ & $\sigma _{\max }$ \\ 
$1024$ & $0.0022 \pm 0.0003$ \\ 
$2048$ & $0.0015 \pm 0.0002$ \\ 
$4096$ & $0.00095 \pm 0.00015$ \\ 
$8192$ & $0.00060 \pm 0.00011$ \\ 
$16384$ & $0.00040 \pm 0.00008$ \\ 
$32768$ & $0.00026 \pm 0.00006$ \\ 
$65536$ & $0.00017 \pm 0.00005$
\end{tabular}
{\rm {,}}  \label{21}
\end{equation}

Curiously, when we decrease the step both the error and the central value of 
$\sigma_{{\rm max }}$ decrease. This can be understood easily, since we take
as value for maximal $\sigma$ in each run the first $\sigma$ for which the
probability after $m_{{\rm max}}$ iterations is too small (smaller than $.7$
in this example), and thus we clearly underestimate it in gross dependence
with the step. We are thus forced to repeat the computation of 
$\sigma_{{\rm max}}$ and $\Delta \sigma_{{\rm max}}$ with smaller and smaller steps, from d$\sigma=10^{-4}$ to d$\sigma=10^{-8}$. We must now fit the dependence of $\sigma _{\max }$ on d$\sigma$ (see Fig. $2$) and extrapolate to d$\sigma=0$.

\begin{figure}[tbp]
\let\picnaturalsize=N
\def\picsize{5.0in}
\def\picfilename{robfig2.eps}
\ifx\nopictures Y\else{\ifx\epsfloaded Y\else\input epsf \fi
\let\epsfloaded=Y
\centerline{\ifx\picnaturalsize N\epsfxsize \picsize\fi \epsfbox{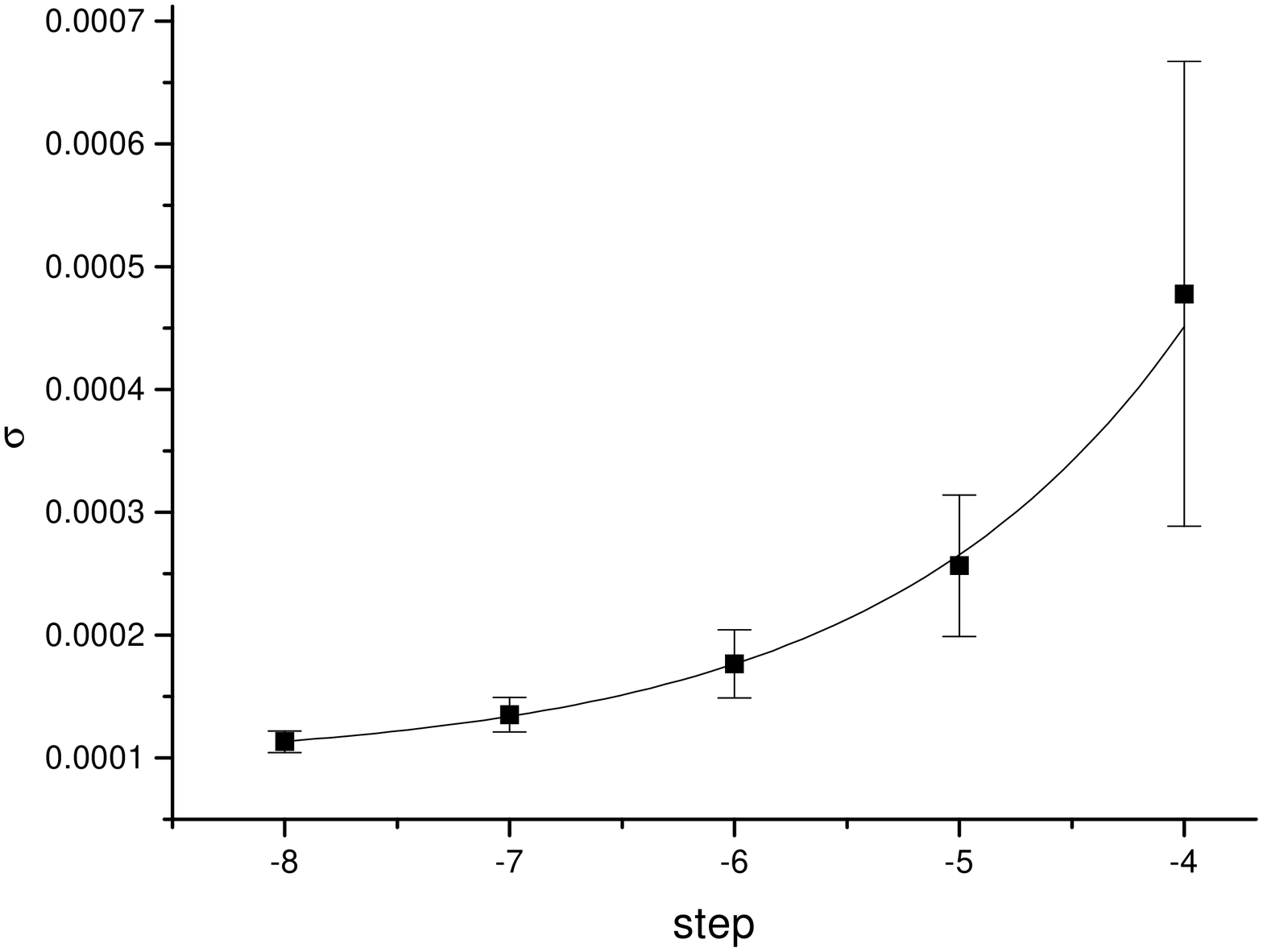}}}\fi
{\small Fig. 2. Plot of $\sigma_{\rm max} $ as a function of $\log {\rm d}\sigma$: the maximum allowed value of noise characterized by $\sigma_{\rm  max} $ before $P_{\rm max}$ $\leq P_{\rm cut}$ depends on the size of the step d$\sigma$ by  which $\sigma $ is increased in the program. This plot is for $N=32768$ and $P_{cut}=0.7$.}
\end{figure}

The generic relation we found is 
\begin{equation}
\sigma _{\max }\left( N,{\rm d}\sigma\right) =\zeta \left( N\right) +\xi
\left( N\right) {\rm d}\sigma^{\alpha }\text{,}  \label{22}
\end{equation}
where $\alpha=0.30\pm 0.06 $ is a true constant. The values of the 
$N$--dependent $\zeta $ and $\xi $ are the following: 
\begin{equation}
\begin{array}{ccc}
N & \zeta & \xi \\ 
1024 & 0.00104 \pm 0.00004 & 0.0240 \pm 0.0046 \\ 
2048 & 0.00065 \pm 0.00001 & 0.0163 \pm 0.0017 \\ 
4096 & 0.00038 \pm 0.00002 & 0.0114 \pm 0.0027 \\ 
8192 & 0.00023 \pm 0.00001 & 0.0086 \pm 0.0020 \\ 
16384 & 0.00015 \pm 0.00001 & 0.0079 \pm 0.0025 \\ 
32768 & 0.00009 \pm 5\times 10^{-6} & 0.0068 \pm 0.0020 \\ 
65536 & 0.00006 \pm 4\times 10^{-6} & 0.0076 \pm 0.0034
\end{array}
\label{23}
\end{equation}
Taking the limit d$\sigma\rightarrow 0$, we obtain the final value of $\sigma _{{\rm max }}$ for each $N$ at this $P_{{\rm cut}}=0.7$: 
\begin{equation}
\begin{tabular}{cc}
$N$ & $\sigma _{\max }$ \\ 
$1024$ & $0.00104\pm 0.00004$ \\ 
$2048$ & $0.00065\pm 0.00001$ \\ 
$4096$ & $0.00038\pm 0.00002$ \\ 
$8192$ & $0.00023\pm 0.00001$ \\ 
$16384$ & $0.00015\pm 0.00001$ \\ 
$32768$ & $0.00009\pm 5\times 10^{-6}$ \\ 
$65536$ & $0.00006\pm 4\times 10^{-6}$ 
\end{tabular}
{\rm {,}}  \label{24}
\end{equation}
The above numbers are very well fit by a straight line (in log~$N$).

From the data (\ref{24}), for this value of  $P_{{\rm cut}}=0.7$, we find finally the relation  
\begin{equation}
\sigma _{\max }\left( N\text{, }P_{cut}=0.7\right) =\alpha \left( P_{{\rm cut}}\right) N^{\phi }  \label{25}
\end{equation}
with $\alpha =0.138\pm 0.012$, and $\phi =-0.704\pm 0.01$.

Similarly, for other values of $P_{{\rm cut}}$ we found : 
\begin{equation}
\begin{array}{ccc}
P_{cut} & \alpha & \phi \\ 
0.5 & 0.158 \pm 0.011 & -0.687 \pm 0.007 \\ 
0.6 & 0.146 \pm 0.010 & -0.691 \pm 0.008 \\ 
0.7 & 0.138 \pm 0.012 & -0.704 \pm 0.010 \\ 
0.8 & 0.083 \pm 0.006 & -0.669 \pm 0.008 \\ 
0.9 & 0.094 \pm 0.001 & -0.724 \pm 0.015
\end{array}
\label{26}
\end{equation}
The value quoted in the text, equation (\ref{xxy}), is an average of these numbers.

To establish equation (\ref{15}), we found for each $P_{\rm cut}$ a table  like (\ref{24}) and then, fixing $N$, we found a good linear fit, equation  (\ref{15}),  with the
following values of  $\gamma(N) $ and $\delta (N)$:

\begin{equation}
\begin{array}{ccc}
N & \gamma & \delta \\ 
1024 & 0.0024 \pm 0.001 & 0.0020 \pm 0.0001 \\ 
2048 & 0.0015 \pm 0.00005 & 0.0013 \pm 0.00005 \\ 
4096 & 0.00092 \pm 0.00003 & 0.00077 \pm 0.00003 \\ 
8192 & 0.00057 \pm 0.00002 & 0.00048 \pm 0.00002 \\ 
16384 & 0.00033 \pm 0.00002 & 0.00026 \pm 0.00002 \\ 
13768 & 0.00021 \pm 0.00002 & 0.00017 \pm 0.00002 \\ 
65536 & 0.00015 \pm 0.00001 & 0.00013 \pm 0.00001
\end{array}
\text{.}  \label{27}
\end{equation}

\end{document}